# Cell contraction induces long-ranged stress stiffening in the extracellular matrix

**Author:** Yu Long Han[1†], Pierre Ronceray[2†], Guoqiang Xu[1], Andrea Malandrino[1,3], Roger Kamm[1,4], Martin Lenz[5], Chase P. Broedersz[6*] and Ming Guo[1*]

*Affiliations:*

1. Department of Mechanical Engineering, Massachusetts Institute of Technology, Cambridge, MA, 02139, USA;

2. Princeton Center for Theoretical Science, Princeton University, Princeton, NJ 08544, USA;

3. Institute for Bioengineering of Catalonia, Barcelona, 08028, Spain;

4. Department of Biological Engineering, Massachusetts Institute of Technology, Cambridge, MA, 02139, USA;

5. Laboratoire de Physique Théorique et Modèles Statistiques (LPTMS), CNRS, Université Paris-Sud, Université Paris-Saclay, 91405 Orsay, France;

6. Arnold-Sommerfeld-Center for Theoretical Physics and Center for NanoScience, Ludwig-Maximilians-Universität München, D-80333 München, Germany

† These authors contributed equally to this work.

[*]**Correspondence**:

Ming Guo

*Massachusetts Institute of Technology, Cambridge, MA 02139, USA.*

*Phone: +1 (617) 324-0136*

*Email: guom@mit.edu*

Chase P. Broedersz

*Ludwig-Maximilians-Universität München, D-80333 München, Germany*

*Phone: +49 (0)89 2180-4514*

*Email: C.Broedersz@lmu.de*

**Conflict of Interest statement:**

The authors declare no conflict of interests.

**Abstract**

Animal cells in tissues are supported by biopolymer matrices, which typically exhibit highly nonlinear mechanical properties. While the linear elasticity of the matrix can significantly impact cell mechanics and functionality, it remains largely unknown how cells, in turn, affect the nonlinear mechanics of their surrounding matrix. Here we show that living contractile cells are able to generate a massive stiffness gradient in three distinct 3D extracellular matrix model systems: collagen, fibrin, and Matrigel. We decipher this remarkable behavior by introducing Nonlinear Stress Inference Microscopy (NSIM), a novel technique to infer stress fields in a 3D matrix from nonlinear microrheology measurement with optical tweezers. Using NSIM and simulations, we reveal a long-ranged propagation of cell-generated stresses resulting from local filament buckling. This slow decay of stress gives rise to the large spatial extent of the observed cell-induced matrix stiffness gradient, which could form a mechanism for mechanical communication between cells.

**Introduction**

Living cells interact mechanically with their three-dimensional (3D) microenvironment. Many basic cell functions, including migration, proliferation, gene expression, and differentiation, depend on how these forces deform and shape the surrounding soft extracellular matrix (ECM) (1-4). In addition, externally imposed forces on the matrix can impact cell behavior, for instance in beating cardiac cells on a two-dimensional (2D) substrate (5-7). Such external forces may be generated by other cells and act as mechanical signals (8-10) leading to emergent collective cell dynamics (11, 12). Nevertheless, it remains unclear how cell-generated forces propagate through the ECM and impact the mechanics of their 3D extracellular environment.

The ECM is comprised of several biopolymers (13), such as collagen or fibrin, which are largely responsible for its mechanical properties. Experiment and theory have shown that biopolymer networks exhibit a highly nonlinear mechanical response (14), involving the entropic elasticity of individual filaments, geometric effects due to fiber bending and buckling, and even collective network effects governed by critical phenomena (15-21). Recent works have indicated that this nonlinear response is highly relevant to cell-ECM interactions (22, 23). Due to a lack of direct characterization of local mechanics in a 3D matrix in the vicinity of a cell, however, the role of elastic nonlinearities in mechanical cell-ECM interactions has remained elusive.

Ideally, cell-ECM interactions would be analyzed by determining the stress field generated by the cell. Unfortunately, standard microscopy techniques do not reveal this information in a straightforward and unambiguous way. Some information about internal network forces can be accessed by adding deformable particles (24) or by creating an interface, for example by laser ablation, and observing the resulting deformation of the system (25, 26). However, obtaining internal stresses with such invasive and destructive approaches requires additional assumptions about the network's local mechanical properties. The same is true of approaches that infer stresses from a combination of microscopy imaging and finite element modelling (23, 27, 28). The intrinsic heterogeneity (29-31) and a highly nonlinear mechanical response (14, 32) of extracellular networks pose a daunting challenge to these techniques (33).

To investigate how living cells mechanically modify their microenvironment, we use microrheology with optical tweezers to directly measure the local nonlinear elastic properties in a 3D ECM network. We observe that remarkably far-reaching stiffening gradients are generated towards the cell in a variety of biopolymer matrices. To investigate this, we introduce a novel model-independent measurement technique termed Nonlinear Stress Inference Microscopy (NSIM), enabling us to determine the stress in a region around the cell and study stress propagation inside a 3D ECM. We use a combination of theory and simulations to demonstrate the ability of NSIM to accurately measure 3D local stress with high spatial resolution. Using NSIM, we show that the observed long-range stiffness gradient around cells results from remarkably long-ranged stresses, which are capable of exciting the matrix's nonlinear response over distances exceeding the size of the cell. Our results demonstrate that cells strongly modify the mechanics of the surrounding ECM, which could be crucial in shaping matrix-mediated cell-cell interactions.

**Cells strongly stiffen their surrounding extracellular matrix by actively contracting**

To study the mechanical interactions between cells and their surrounding ECM, we culture MDA-MB-231 cells in a 1.5 mg/mL reconstituted 3D collagen network. The network is infused with 4.5 µm diameter latex beads, large enough to prevent slippage through the mesh. The cells spread and start contracting the surrounding network within 4 hours (Fig. 1a). We probe the local micromechanics of the matrix using optical tweezers to pull these beads away from the cell at a constant speed of 1 µm/s (Figs. 1b-d and Supplementary Fig. 1). This low speed ensures that the viscous drag on the bead from the background fluid is negligible compared to the network's restoring force, and at this speed the mechanical response of the matrix is rate-independent, fully reversible and therefore predominately elastic (Supplementary Fig. 2). Thus, this protocol enables us to obtain the local force-displacement relationship $F(x)$ that characterizes the micromechanics of the matrix.

By probing a bead located far from the cell (>200 µm), we determine the intrinsic response of the collagen matrix. The resulting force-displacement relationship is shown by the black line in Fig. 1e. The nonlinear differential stiffness $k_{nl}(F)=dF/dx$, defined as the slope of this force displacement curve, increases with applied force $F$, revealing a strong force-stiffening behavior. This is reminiscent of the well-characterized stress-stiffening behavior measured at large scales using macrorheology on collagen gels (14, 32).

Interestingly, the matrix becomes substantially stiffer closer to the cell (Fig. 1e). Indeed, the local linear stiffness $k_{lin}$ of the matrix, defined as the small force limit of $k_{nl}(F)$, is two orders of magnitude larger near the cell than at a remote location (Fig. 1f, red squares). This observation in the bulk of the network is consistent with prior 2D experiments showing cell-induced stiffening of the surface of a collagen matrix with a cell migrating on top (22), as well as with simulations (23).

This dramatic stiffness gradient in the vicinity of the cell originates from the active forces it exerts. Indeed, inhibiting cell contractility using 2 µM Cytochalasin D results in a strong attenuation of the cell-induced stiffening (Fig. 1f, blue circles). The weak residual gradient is well explained by increased ECM density near the cell, under the assumption that the matrix rigidity scales as the square of the collagen concentration (*c*) (34) (Fig. 1f, gray diamonds).

Indeed, by estimating $c$ using confocal reflection microscopy, we determine that the enhanced matrix density near the cell can account for a stiffening of up to a factor ~3 (Fig. 1f, gray diamonds). Finally, we rule out the effect of the passive rigidity of the cell on the matrix stiffness, which is very short-ranged in 3D (31). Thus, active forces exerted by the cell result in an extended stiffened region in the 3D matrix, reflecting the presence of a stress field decaying away from the cell with stress values sufficiently large to excite the nonlinear response of the collagen network, as illustrated in Fig. 1d.

**Nonlinear Stress Inference Microscopy**

To study the cell-induced stress fields, we use the network's nonlinear microrheological response to our advantage and infer local stress values from our stiffness measurements. The nonlinear stiffening evidenced in Fig. 1e originates from two contributions: the force $F$ exerted by the optical tweezers acting on the bead, and the local stress $\sigma_{loc}$ induced by the cell. This similar influence of force and stress suggests that we may be able to extract $\sigma_{loc}$ at a specific distance from the cell by comparing the corresponding force-displacement relationship to the remote measurement at which $\sigma_{loc}$ is negligible. This comparison is confounded, however, because of force and stress being fundamentally different quantities: beyond having different dimensions, force transforms as an axial vector under spatial symmetry operations, while stress is a rank two tensor. This has an essential implication for the difference in the nonlinear response due to a force as opposed to a stress: The local stiffness should be invariant under reversal of the force vector, $F$ to $-F$, while reversal of the stress tensor $\sigma$ to $-\sigma$ exchanges compression and tension, which can have a qualitatively different effect on the nonlinear mechanical response.

Despite these differences, here we show that a correspondence between force and stress controlled stiffening can be established in the strongly nonlinear regime. First consider a simple 1D system of nonlinear springs representing the network surrounding a bead in a geometry with fixed network stress $\sigma$ (Fig. 2a) and one with fixed tweezer force $F$ (Fig. 2b). For nonlinear springs that stiffen under tension and soften under compression --- a generic characteristic of biopolymers (14, 18)--- we find that the functional form of the stiffness curves actually become similar at large $\sigma_{loc}$ and $F$, despite being qualitatively different in the weakly nonlinear regime. Indeed, the tensed spring in Fig. 2b dominates the differential stiffness experienced by the bead in the strongly nonlinear regime, rendering this case similar to the stress-controlled geometry, where the mechanical response is equally shared by two similarly tensed bonds (Fig. 2a). This quantitative similarity between the $k_{\text{lin}}$ vs. $\sigma_{\text{loc}}$ and $k_{\text{nl}}$ vs. $F$ curves in the strongly nonlinear regime enables us to use the latter, which we measure by nonlinear microrheology, as a "dictionary" to infer local stresses.

This intuitive similarity between the force and stress controlled geometries in the nonlinear regime becomes mathematically exact when the springs' differential stiffness has a power-law dependence on tension, as widely observed for biopolymer networks (32, 35) (SI section 4). Specifically, from a measurement of $k_{\text{lin}}$ in a network with an unknown local stress $\sigma_{\text{loc}}$, we can obtain an effective force $F_{\text{eff}}$ defined such that $k_{\text{nl}}(F_{\text{eff}}) = k_{\text{lin}}(\sigma_{\text{loc}})$, and this effective force is directly proportional to the local stress

$$\sigma_{\text{loc}} \approx a\, F_{\text{eff}} \tag{1}$$

provided large local stresses, such that $k_{\text{lin}} \gg k_0$, where $k_0$ is the linear stiffness of the unstressed network. We determine the proportionality factor $a$ by assuming that nonlinearity sets in at a similar stress $\sigma^*$ at a macro and microscopic level. In practice, we adjust $a$ to match the low- and high-stress asymptotes, in a log-log plot, of the macroscopic differential shear modulus $K(\sigma_{\text{macro}})$ to those of the microrheology curve $k_{\text{nl}}(F)$ (Figs. 2c-e, Supplementary Figs. 5 and 8). Together with Eq. (1), this provides a procedure to infer stresses from nonlinear microrheology, which we term Nonlinear Stress Inference Microscopy (NSIM).

To demonstrate the validity and accuracy of NSIM, we perform simulations of the experimental scenario in Fig. 1. We embed a contractile cell in a disordered 3D network of fibers with power-law stiffening (SI Section 2). We model the cell as a rigid ellipsoidal body that contracts along its long axis, inducing strong stiffening in an extended conic region as depicted in Fig. 3a. We then simulate a microrheology experiment by applying a force on a selection of network nodes to obtain a local force-displacement curves at various distances $r$ along the contraction direction of the cell (Fig. 3b). From this, we determine the linear stiffness $k_{\text{lin}}$ of the network as a function of $r$ (Fig. 3c), which exhibits a dependence similar to the experimental measurements shown in Fig. 1f. We further confirm that this dependence vanishes in the direction perpendicular to contraction and in the absence of an active contractile force, as in experiments (Fig. 1f). We infer the local stress field from these linear stiffnesses using NSIM, as shown in Fig. 3d. We find excellent agreement with the "true" local stress in the strong stiffening regime even when mechanical disorder gives rise to fluctuations in the stress field (Fig. 3e and SI Section 3), thereby validating NSIM as a quantitative method to capture the spatial stress distribution around a contractile cell in a disordered 3D fiber matrix (Figs. 3d-f).

**Long-ranged stress propagation leads to extended stiffness gradients around cells**

To unravel the mechanical origins of the far-reaching cell-induced stiffness gradient in collagen (Figs. 1,4a), we use NSIM to experimentally infer the local stresses $\sigma_{loc}(r)$ around a cell inside the matrix. The inferred stress decays consistently with a power-law $\sigma_{loc} \sim r^{-2}$ over more than two decades (Fig. 4b), in contrast with the power law $\sigma_{loc} \sim r^{-3}$ expected in a linear material (36). This measured slow stress decay is consistent with both our simulations (Fig. 3f) and previous theoretical predictions for nonlinear fiber networks (37-39). Deviations from linear elasticity have previously been reported in experiments for the deformation fields (23, 27), but the implications for the stress-field have remained unclear (40). Here, we establish a direct experimental measurement of the long-ranged transmission of cell generated stresses in fiber matrices.

Conceptually, this increased range of stresses in fibrous materials found in simulations results from their asymmetric response to tension and compression: Fibers stiffen under tension and soften due to buckling under compression (18, 41). Simply speaking, the matrix around a strong contractile cell effectively behaves as a network of ropes, where only tensile forces are transmitted, unimpeded by orthoradial compressive resistance. Hence the total contractile force exerted by the cell is conserved with distance, and the decay of radial stress simply reflects this force spreading over an increasing surface area (37). This buckling-based mechanism is supported by observations with confocal reflection microscopy of a larger amount of highly curved collagen filaments in the vicinity of a contractile cell, as compared to the case where contraction is inhibited with Cytochalasin D (Fig. 4e-f).

To explore the generality of our observations in collagen, we perform the same nonlinear microrheological experiments with MDA-MB-231 cells in a 2.5 mg/mL Matrigel (Fig. 4a, green circles), a blend of biopolymers more complex than pure collagen (42), and for human umbilical vein endothelial cells (HUVEC) in a 3.0 mg/mL fibrin gel (light blue triangles in Fig. 4a). In both cases, we find that cells are capable of generating large extended stiffness gradients along the cell's contraction direction (Fig. 4a). Using NSIM, we find that the slow stress decay consistent with rope-like force transmission ($\sigma_{loc} \sim r^{-2}$) is observed in all three cases, despite significant variability in the absolute magnitude of the stresses (solid lines in Fig. 4b). These results highlight the wide applicability of NSIM, and demonstrate the generality of long-ranged stress fields generated by the cell. This long-ranged stress transmission is intimately tied to the ability of the cell to enhance the linear stiffness $k_{lin}$ over an extended region of the ECM (Figs. 1f, 4a): Slowly decaying stresses induced by the cell remain large enough to excite the nonlinear elastic response of the matrix over a distance exceeding the cell size.

Cells not only actively modify the linear stiffness of their 3D matrix environment (Fig. 4a), but also the nonlinear mechanical response. To reveal how cell stress and probe forces combine to stiffen the surrounding network, we measure the full nonlinear microrheological response of the network in the vicinity of the cell in all three types of ECM model systems, as shown in Fig. 4c. Indeed, the nonlinear $k_{nl}$ vs. $F$ curves measured at different distances from the cell are clearly separated, indicating a significant contribution of cell generated stress on the nonlinear mechanical response. This contribution could be through nonlinear network elastic stiffening or network plastic deformation. We note that our stress inference is largely independent of the specifics of the ECM's nonlinear response, but does assume a predominantly elastic response to the forces generated by the cell. Indeed, significant plastic deformations could imply that the ECM's nonlinear response is systematically modified as a function of the distance from the cell. In the absence of plastic deformations, we expect that further stiffening a prestressed matrix by a large tweezer force would result in a nonlinear response that is functionally similar for all levels of cell stress. To test this, we plot all nonlinear stiffening curves as a function of the combined force $F+F_{eff}$, where the effective force $F_{eff} \propto \sigma_{loc}$ is determined as in Figs 2a-b (Supplementary Table 1). Remarkably, we find that the data taken at different distances to the cell collapse in any network composition onto a smooth master curve (Fig. 4d). The large cell-generated stress thus locally drives the ECM into an elastic nonlinear regime, which can be further extended by the probe force we apply with optical tweezers.

Cell contraction induces a local stiffening of the surrounding ECM by several orders of magnitude with a long-range decay over tens of micrometers. Here we infer the stresses responsible for this stiffening using NSIM, a conceptually new inference technique that does not require the materials' constitutive relationship of the stress field in terms of the strain field nor a reference undeformed state. Due to its simplicity and insensitivity to the detailed material's properties, NSIM could be used in various conditions, including embryo or tumor development. The stresses inferred using this technique far from the cell are consistent with prior measurements (27). Close to the cell, high stiffening renders the technique most accurate, and corresponds to stresses of the order of 200 Pa, significantly larger than previously reported (27). These cell-induced stresses decay much more slowly than in a linear continuum material due to buckling in fiber networks, resulting in far-reaching stiffness gradients as high as 50 kPa/mm over several cell diameters. Other cells in the surrounding matrix could sense and respond to such large gradients, suggesting that cell-induced ECM stiffening could mediate inter-cell mechanical communication and collective durotaxis. These observations highlight the critical role of nonlinear matrix mechanics not only in shaping cell-ECM interactions (8, 43), but also for matrix-mediated interactions between cells.

**Methods**

**Cell culture and matrix preparation.** Cells are maintained under 37 °C, 5% CO2 and 95% humidity. MDA-MB-231 cells were cultured in DMEM with 10% FBS, 1% penicillin and streptomycin. Human Umbilical Vein Endothelial Cells (HUVEC) (Lonza) were cultured on collagen I-coated asks in EGM-2 growth medium (Lonza) and used between passages 6–8. To prepare the collagen gel, 800 µL type I bovine collagen solution (3.0 mg/mL, PureCol, Advanced BioMatrix) was mixed with 100 µL PBS (10X). We adjusted the solution to pH 7.2 with ~70 µL 0.1 M NaOH. The collagen solution is then mixed with PBS (1X) to reach a final collagen concentration of 1.5 mg/mL, and polymerized in the cell culture incubator for 30 min. To prepare the fibrin gel, fibrinogen from bovine plasma (F8630, Sigma) was dissolved in PBS at 6 mg/mL. Thrombin (T4648, Sigma) was dissolved at 2U/mL in PBS (for experiments without cells) or in EGM-2 (for experiments with cells). Then we mixed thrombin and fibrinogen at 1:1 volume ratio and polymerized it in the cell culture incubator for 15 min. For Matrigel preparation, the basement membrane matrix (10 mg/mL, Corning) was diluted to 2.5 mg/mL with DMEM and polymerized in the cell culture incubator for 30 min. For all cell-loaded gels, cell and bead suspensions were added to the gel solution before polymerization, with a cell density around $10^4$/mL, and all measurements were conducted 12 hours after polymerization. To inhibit contractility of MDA-MB-231 cells, we disrupted filamentous actin structures using 2-μM Cytochalasin D (PHZ1063, Invitrogen) for 30 min.

**Optical tweezer measurements.** We used a Thorlabs optical tweezers system to perform all measurements. Briefly, to optically trap a bead (4.5-μm carboxylate microspheres, Polyscience) that is embedded and confined in a 3D biopolymer network, the laser beam (5W, 1064 nm) is tightly focused through a series of Keplerian beam expanders and a high numerical aperture objective (100 x 1.4, oil, Leica). A high-resolution quadrant detector was used for position detection. The linear region of the detector and the trap stiffness (0.04 pN/nm) were calibrated with the same bead in pure cell culture media by using an active power-spectrum method and equipartition theorem (44). To manipulate the trapped bead, a high-resolution piezo stage (P-545, PI nano™) was moved at a constant velocity of 1 μm/s, and the relative distance between laser and bead) was recorded, from which local force-displacement curves inside the matrix were determined (45) (see Supplementary Information for details).

**Bulk rheology.** We performed bulk rheology measurements on a DHR-3 rheometer (TA Instruments) using a plate-plate geometry, with a 40-mm glass disk as the top plate and a 60-mm Petri dish as the bottom plate with a gap of 500 μm. All gels were formed in the gap at 37 °C and were sealed by mineral oil to avoid evaporation. The polymerization process was monitored by strain oscillations with a strain amplitude of 0.005 at a frequency of 1 rad/s. After polymerization, a strain ramp was applied to the gel at a rate of 0.01/s, and the resulting stresses were measured.

**Theoretical modelling and simulations.** Numerical simulations presented in Fig. 3 are performed using a model of nonlinear springs (force-extension relation $f(x) = \exp(\mu x) - 1$), with regular removal of springs to introduce disorder in the network, while ensuring a fiber length $L_f$ =10, in a spherical system of radius $R$=25.5. The contractile cell is a rigid ellipsoidal body of size 14.2 × 2.8 × 2.8, with force and torque balance, contracted by 50% along its long axis. The surrounding network is flexibly clamped at the surface of the cell and at the boundary of the system. Mechanical equilibrium is attained by minimization of the energy using the BFGS algorithm. Further details are provided in SI Sections 2 and 3.

**Imaging of collagen networks and image analysis.** The 3D collagen networks near a contracting cell were imaged with confocal reflection microscopy using a 63x 1.2NA water objective (Leica SP8). To determine the boundary of the cell, the cytoplasm was stained with cell tracker green (C7025, Thermofisher) and imaged at the same time under confocal microscope. To capture the fiber buckling process, we imaged the cell and its surrounding 3D fiber networks at a 5 min interval for 4 hr at 37 ºC and with 5% $CO_2$. To analyze the curvature of single collagen fiber, we manually selected 20 points on each individual collagen fiber; the fiber outline was determined by cubic spline interpolation, from which the average curvature of the fiber was calculated.

## Acknowledgements

The authors thank Anna Posfai for useful comments. This work was supported by the National Cancer Institute grant number 1U01CA202123 (to M.G.), the German Excellence Initiative via the program 'NanoSystems Initiative Munich' (NIM) (to C.P.B.) and the Deutsche Forschungsgemeinschaft (DFG) via project B12 within the SFB-1032 (to C.P.B.), a PCTS fellowship (to P.R.), and a MISTI-Germany seed fund (M.G. and C.P.B.). M. G. would also like to acknowledge the support from the Department of Mechanical Engineering at MIT. A.M. is supported by EU's Seventh Framework Programme for Research (FP7, 625500). This work was performed in part at the Aspen Center for Physics, which is supported by National Science Foundation grant PHY-1607611.

# Figures

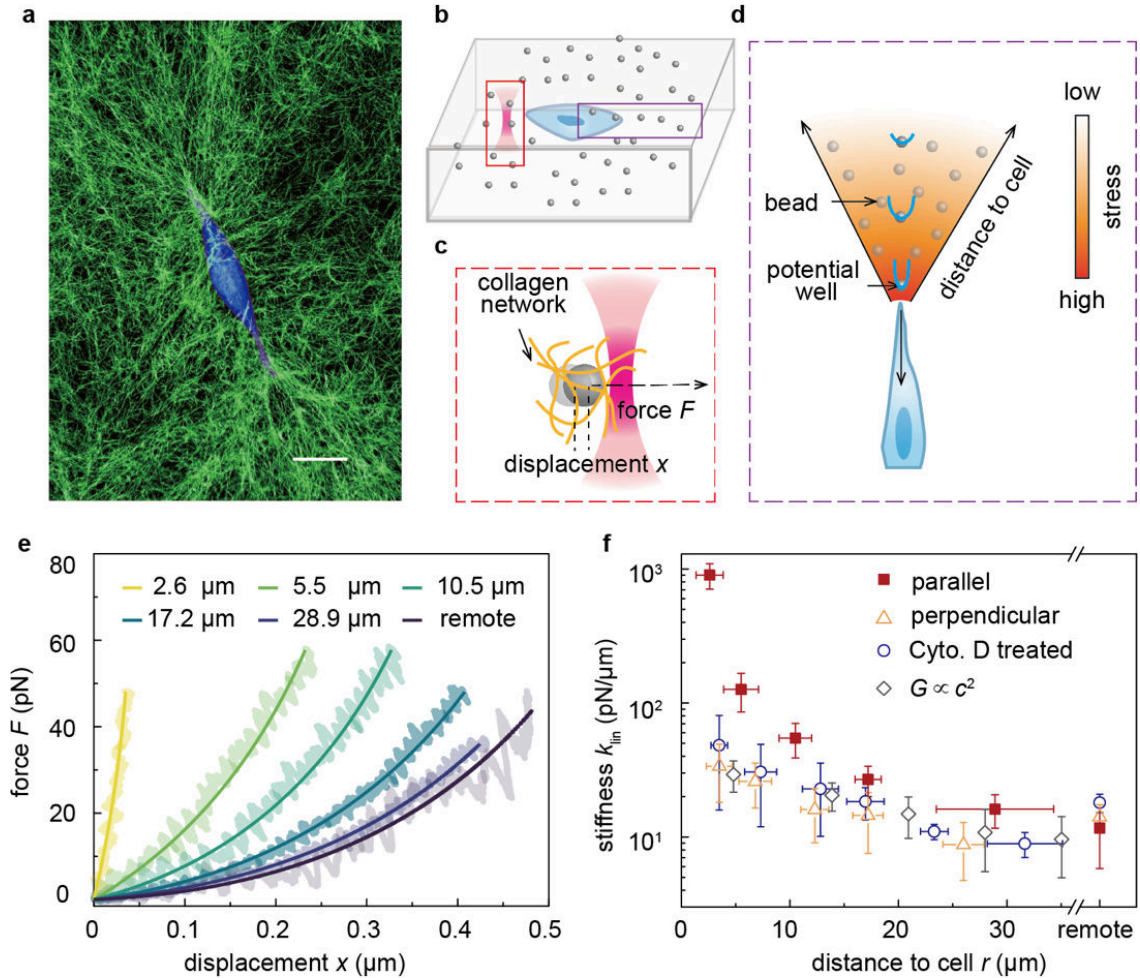

**Figure 1. Far-reaching stiffness gradient of extracellular matrix caused by a single contracting cell in a three-dimensional collagen network. a,** Image of a MDA-MB-231 cell (blue) in a 3D collagen network (green). Scale bar, 10 µm. **b-d**, Schematics illustrating the force-displacement measurement with laser tweezers and the relation between matrix stiffening (blue potential wells) and the cell-generated stress field in the cell contraction direction. **e**, Local force-displacement curves, showing the local nonlinear stiffening response in the collagen network. Different colors represent measurements at various distances from the cell along the contraction direction. **f**, Quantification of the linear stiffness $k_{\text{lin}}$ of the local 3D matrix as a function of distance to the cell $r$. Red and yellow symbols represent measurements along and perpendicular to the main contraction direction, respectively. Blue symbols are measured with cell contraction inhibited by Cytochalasin D treatment. Gray diamonds indicate the stiffness expected solely from the increased collagen concentration $c$. Here, "remote" stands for the locations that are far away from the cell (>200 µm), where the matrix's response is not affected by cell contraction. Error bars represent standard deviation ($n$=15).

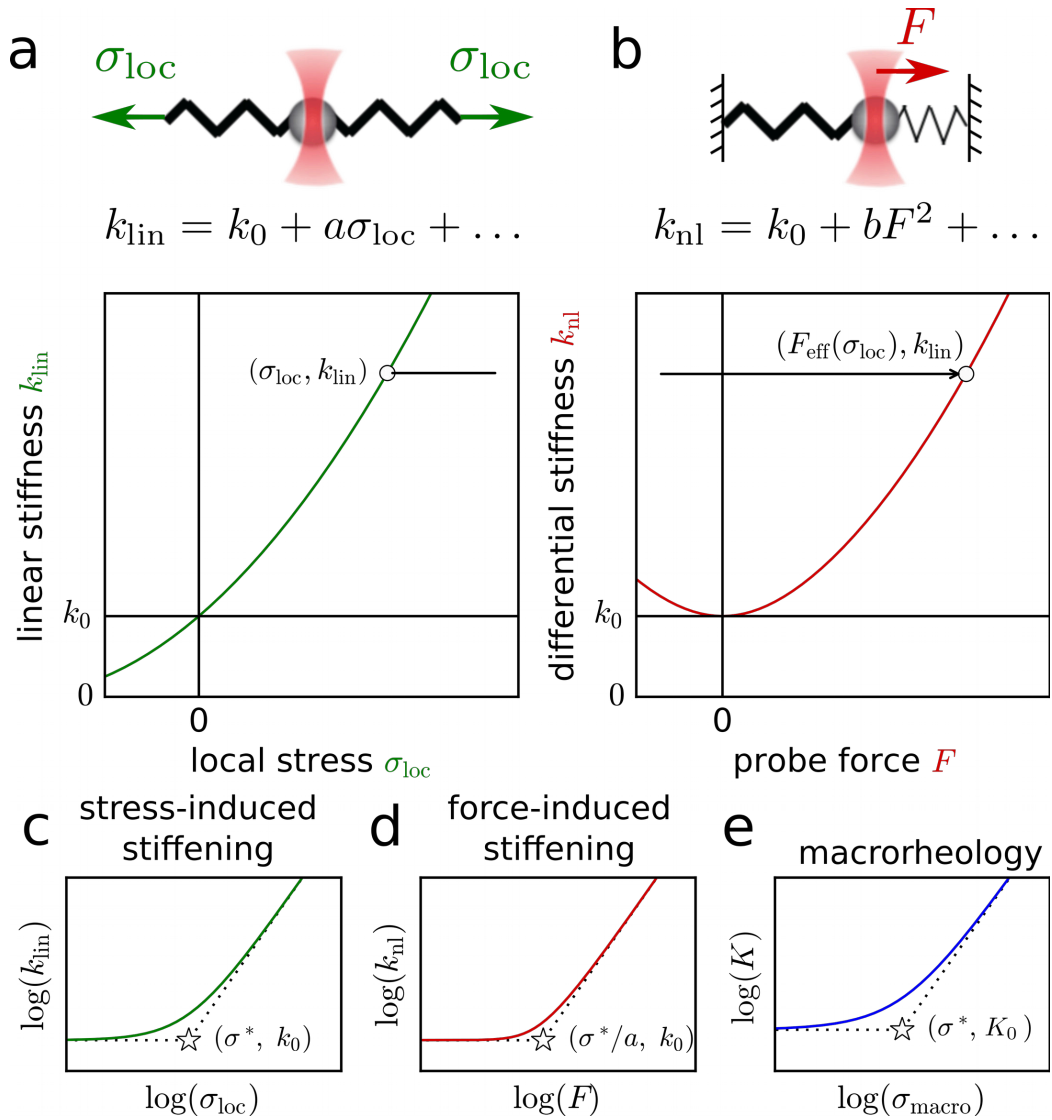

**Figure 2. Nonlinear elastic responses can be used to infer cell-induced local stresses.** **a,** 1D system of nonlinear springs in a stress-controlled geometry with local stress $\sigma_{\text{loc}}$. **b,** force controlled geometry with force $F$ applied to the central bead, together with an expansion of stiffness dictated by symmetry properties of the two scenarios and a schematic of the nonlinear response. The linear stiffness, $k_{\text{lin}}$, of the system in **a** can be measured by applying a small perturbation to the central bead, while the nonlinear stiffness, $k_{\text{nl}}$, is defined as the derivative of the force-displacement curve of the central bead in **b**. The springs represent the surrounding network. **c,** Schematics of linear microrheological stiffness as a function of the local stress in the stress-controlled geometry on a logarithmic scale. **d,** Nonlinear microrheological stiffness for the force-controlled geometry. **e,** Differential shear modulus, $K$, as a function of applied shear stress $\sigma_{\text{macro}}$ as in a macroscopic rheology experiment. Our inference technique exploits a correspondence between the stress-controlled and force controlled geometries in the strongly nonlinear regime.

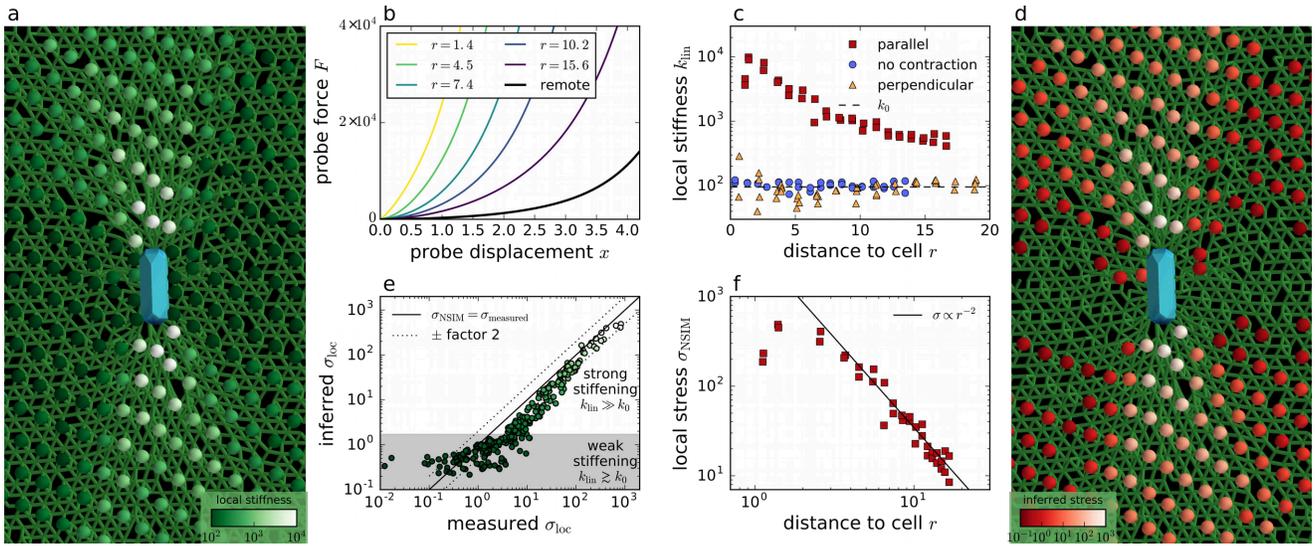

**Figure 3. 3D Simulations of cell-generated stress fields inducing nonlinear network response and validations of NSIM.** **a**, Simulated rigid ellipsoidal cell contracting inside a 3D nonlinear fiber network (in green). The linear stiffness $k_{\text{lin}}$ is depicted by the spheres in a green-white logarithmic color gradient. **b**, Local force-displacement curves, showing the local nonlinear stiffening response in the simulated network. Different colors represent measurements at various distances from the cell along the contraction direction. **c**, Local linear stiffness $k_{\text{lin}}$ of the 3D matrix as a function of distance to the cell $r$ from simulations. Red and yellow symbols represent data parallel and perpendicular to the main contraction direction, respectively. Blue symbols correspond to a non-contracting rigid cell. **d**, The inferred stress depicted by spheres in a red-white logarithmic color gradient in the same simulation as in **a**. Absent points along the direction perpendicular to the cell's contraction axis correspond to soft compressed regions where the local stiffness is smaller than $k_0$, precluding the use of NSIM. **e**, Inferred stresses from simulated data in **b** using NSIM versus direct numerically determined stress, demonstrating that NSIM allows to correctly infer stresses within a factor of 2 in the nonlinear regime. **f,** Local stress obtained from NSIM as a function of distance.

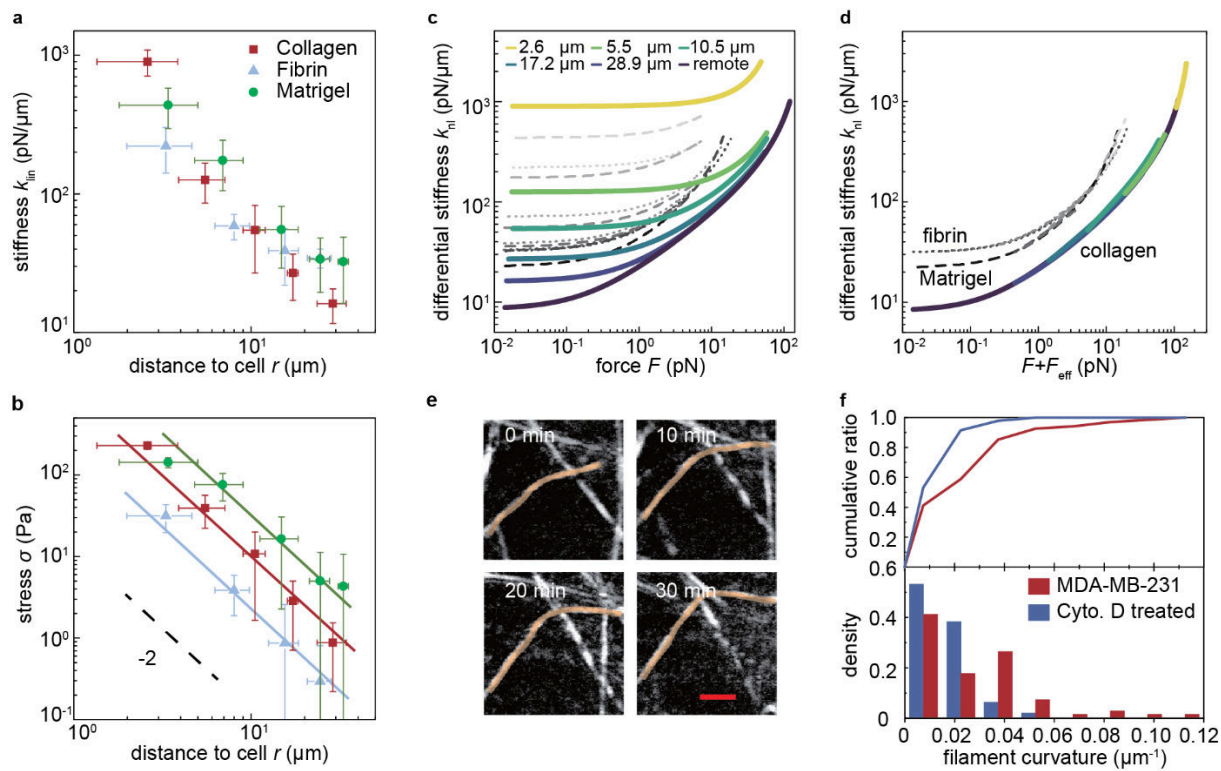

**Figure 4. Nonlinear matrix stiffening and long-ranged stress propagation in various 3D biopolymer networks. a**, Local linear stiffness $k_{lin}$ is plotted against the distance to the cell $r$ along its principal contraction direction in collagen (red square), fibrin (blue triangle) and Matrigel (green circle). All three different ECM model systems exhibit a strong cell induced stiffening gradient. **b**, The stress field $\sigma$ generated by the cell determined using NSIM is shown as a function of distance to the cell $r$, and the dashed line indicates a slope of -2. **c**, Local nonlinear differential stiffness $k_{nl}$ is plotted against the applied probe force $F$ for all three ECM model systems. **d**, Collapse of the date from panel $c$ onto a master curve in each respective matrix obtained by plotting $k_{nl}$ as a function of combined local force $F+F_{eff}$, where the $F_{eff}$ is determined using NSIM. **e**, Time-lapse imaging shows the buckling process of a single fiber around a contracting cell. The fiber undergoing buckling is highlighted in yellow. Scale bar, 2 µm. **f**, Fiber curvature distributions (bottom) and the cumulative probability (top) near the cell, within a 60 µm distance along the principle cell contraction direction, before and after Cytochalasin D treatment. Error bars in **a** and **b** represent standard deviation ($n$=15).

# Supplementary Information for

# Cell contraction induces long-ranged stress stiffening in the extracellular matrix

**By** Yu Long Han, Pierre Ronceray, Guoqiang Xu, Andrea Malandrino, Roger Kamm, Martin Lenz, Chase P. Broedersz and Ming Guo

## 1 Experimental methods and supplementary experimental results.

### 1.1 Data analysis for active microrheology and measurements of collagen micromechanics

To manipulate a trapped bead, a high-resolution piezo stage is displaced at a constant velocity, $v_{stage}$ = 1 µm/s. The distance between the laser and the bead is recorded using a quadrant photodiode detector (QPD) as a voltage signal, as shown by the black curve in Supplementary Figure 1a. This data is fitted with a quadratic function, $V(t) = at^2 + bt$ (red curve), which is used in the following analysis to calculate the differential stiffness, $k_{nl}$. The voltage from the QPD is converted into the distance between the laser and the bead, $D_{laser}$, through a proportionality factor β, i.e. $D_{laser}(t) = \beta V(t)$, as shown by the black curve in Supplementary Figure 1b. The bead displacement relative to the stage is then calculated using $x(t) = v_{stage}*t - D_{laser}(t)$, as depicted by the red curve in Supplementary Figure 1b. The force on the bead is proportional to the distance between the laser and the bead, $F(t) = k*D_{laser}(t)$, where $k$ is the trap stiffness.

The force-displacement relationship for a bead inside a collagen gel is then obtained as shown in Supplementary Figure 1c. The slope of this curve, which represents the effective stiffness of the gel, increases significantly as the applied force increases, suggesting a strong nonlinear stiffening effect in the collagen gel at the microscale. Indeed, the stiffness of the gel increases by two orders of magnitude as the applied force is increased, as indicated by the relationship between differential stiffness, $k_{nl}$, and applied force shown in Supplementary Figure 1d.

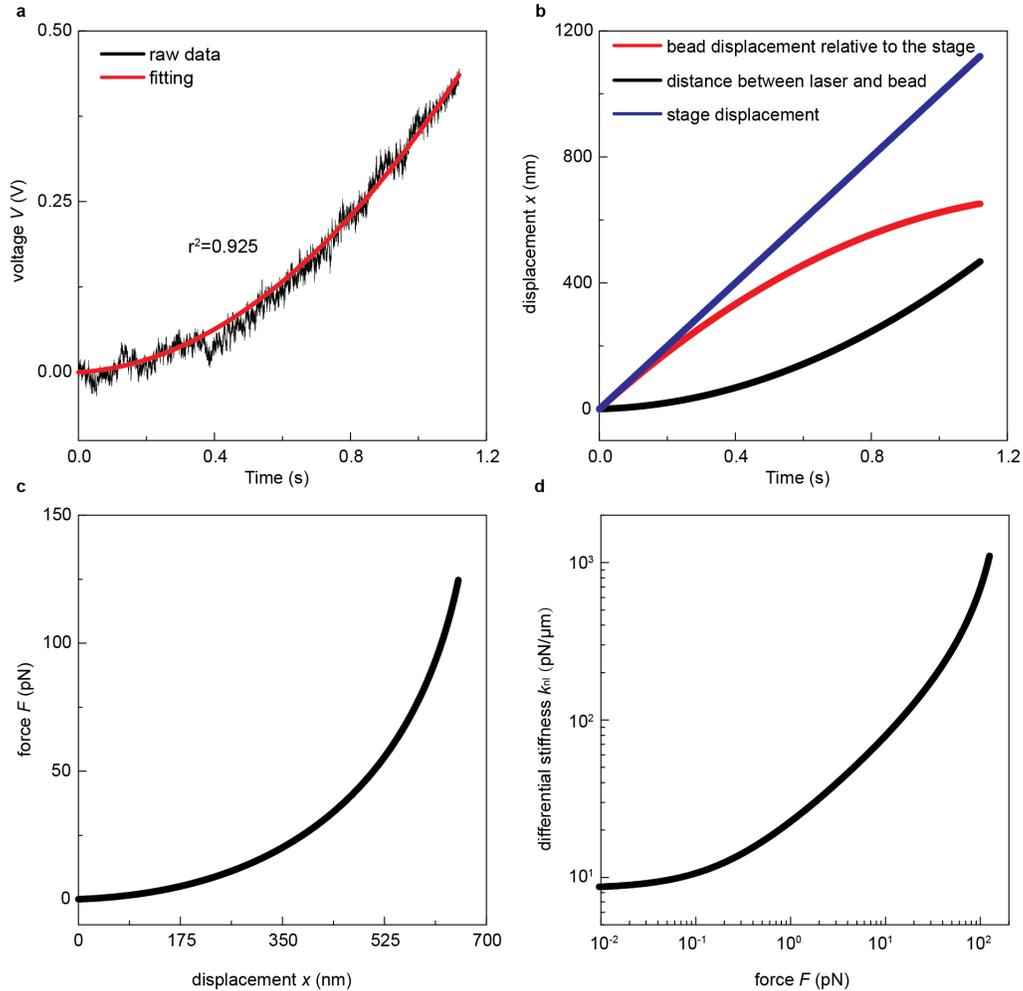

**Supplementary Figure 1:** Data analysis for active microrheology. **a**, Voltage reading from the quadrant photodiode detector (QPD) during mechanical testing, which reflects the relative position between laser trap and particle. The black curve shows the raw data, and the red curve shows the quadratic fit. **b**, Position information of laser, bead, and stage. Stage displacement stands for the displacement that the stage moves. **c**, The force-displacement curve obtained in 1.5 mg/mL collagen I sample far away from the cell. **d**, Force-differential stiffness curve obtained in 1.5 mg/mL collagen I sample far away from the cell.

## 1.2 Rate dependence and reversibility of micromechanical response

To investigate whether the mechanical response of the collagen matrix is rate dependent, we drag a 4.5-μm-diameter bead using optical tweezers to measure the force-displacement relationship at three different locations in a 1.5 mg/mL collagen I gel; at each location, we drag the bead three times with three different speeds, 0.5, 1, 1.5 μm/s, respectively, to measure the collagen micromechanics at different effective strain rates. We find that the resultant force-displacement relationships obtained at varying speeds are very close to each other (Supplementary Figure 2a), suggesting that the micromechanics of the collagen matrix is rate-independent within the range of loading speeds in our study. Furthermore, we have also confirmed that the nonlinear force-displacement relationship measured in the collagen sample is fully reversible with negligible plastic effects, as similar force-displacement curves are obtained from repeated cyclic loading using the same bead, as shown in Supplementary Figure 2b.

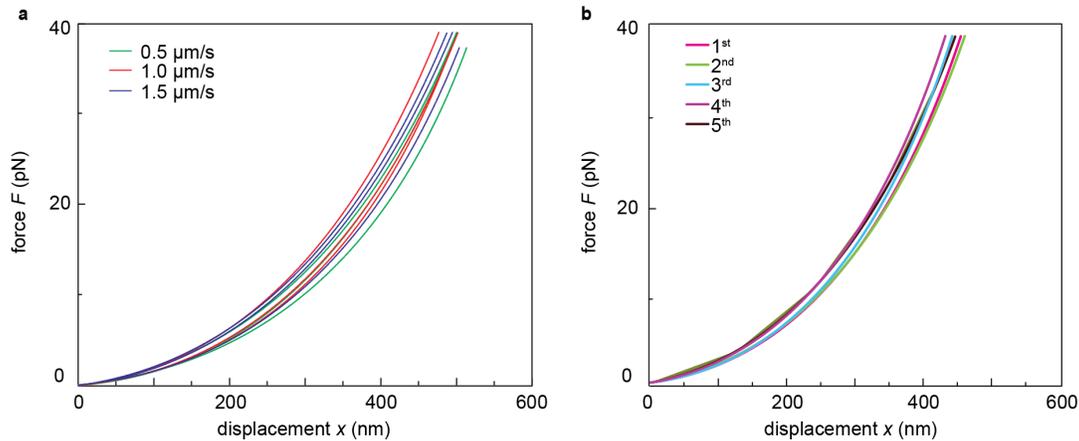

**Supplementary Figure 2:** Micromechanics of collagen matrix measured with optical tweezers. **a**, Force-displacement relationship measured using optical tweezers at different loading rates at 3 different locations in a 1.5 mg/mL collagen gel. At each location, 3 different speeds are tested: 0.5 μm/s (green), 1.0 μm/s (red), and 1.5 μm/s (blue). This result indicates that the mechanical response of the local matrix is largely rate-independent. **b**, Force-displacement relationship of five subsequent loading cycles at the same location in a 1.5 mg/mL collagen gel. This result indicates that the collagen matrix is predominately elastic in our study.

To check the uncertainty and local variability of our measurements, 10 independent measurements in 1.5 mg/mL collagen gel far from a cell are performed, as shown in Supplementary Figure 3a. Interestingly, while the force-displacement curves can differ significantly from location to location, the resultant relationship between the differential stiffness and the applied force are similar, especially in the nonlinear regime, as shown in Supplementary Figure 3b.

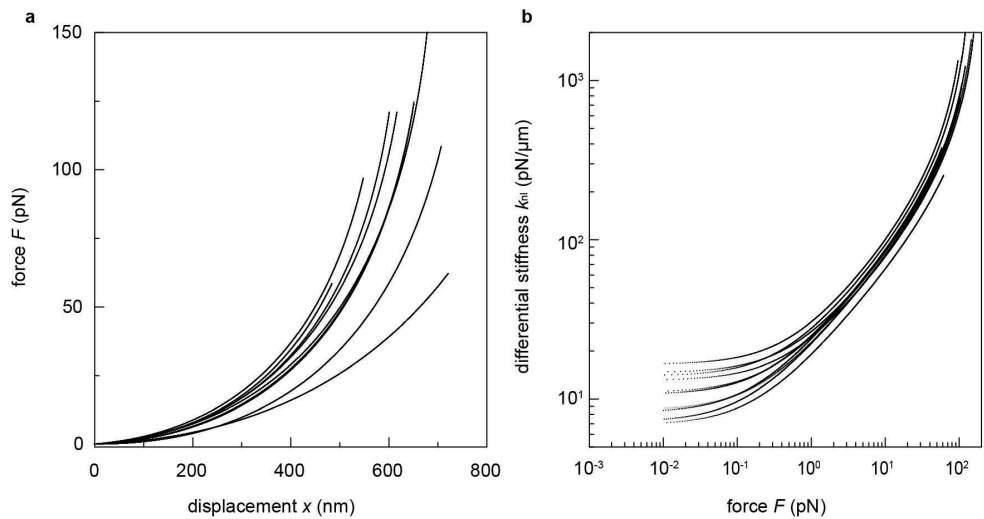

**Supplementary Figure 3:** Optical tweezers microrheology measurements in collagen samples. **a**, Ten independently measured force-displacement curves. **b**, Corresponding differential stiffness-force ($k_{nl}$ vs. $F$) curves. The result indicates that although the collagen network is heterogeneous in linear stiffness, the nonlinear mechanics share a similar trend and collapse together.

Similar behavior is also observed in the cytochalasin D treated sample. We measure the nonlinear micromechanics at different distances to the cell in a 1.5 mg/mL collagen gel, with cell contractility inhibited by treatment of cytochalasin D, and plot the stiffness-force ($k_{nl}$ vs. $F$) curves. We find that although the linear stiffness $k_{lin}$ slightly increases closer to the cell, consistent with the observed increase in collagen density (Fig. 1f), the nonlinear mechanics at all distances approximately overlap for high forces, as shown in Supplementary Figure 4. This result indicates that the matrix density does not significantly affect the nonlinear mechanical properties of the collagen gel, consistent with previous work [32].

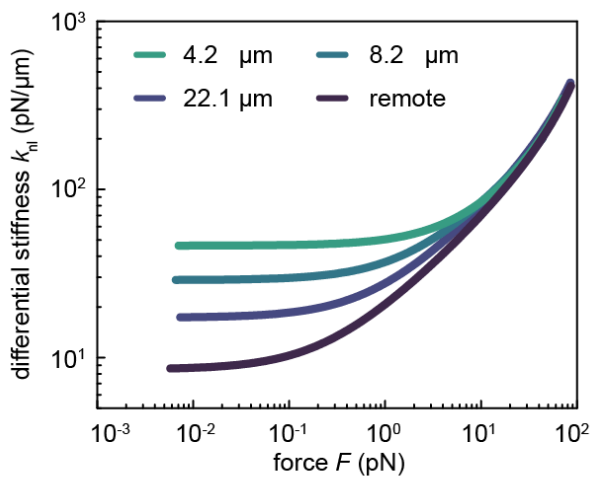

**Supplementary Figure 4**: Local nonlinear differential stiffness $k_{nl}$ is plotted against the applied probe force $F$ in a collagen matrix near a cytochalasin D treated cell. The difference in the linear stiffness $k_{lin}$ is consistent with a change in fiber density, but all curves at different distances to the cell appear to asymptotically converge to the same response in the nonlinear regime, consistent with previous work [32], indicating that the matrix density does not affect the nonlinear mechanical properties of the collagen gel.

### 1.3 Conversion from local force on the bead to local matrix stress

To use NSIM, we require a conversion between local force on the bead to local matrix stress (See Eq. 1 in the main text). This conversion is established by comparing nonlinear microrheology data with nonlinear macrorheology data. Specifically, we first shift the microrheology data ($k_{nl}$ vs. $F$) in the vertical direction to align these data to the linear regime of the macrorheology data ($K$ vs. $\sigma$), as shown in Supplementary Figure 5a. Subsequently, we shift the microrheology data in the horizontal direction to align these data with the nonlinear regime of the macrorheology data (Supplementary Figure 5a); the shift factor in the horizontal direction involves a multiplication factor, which defines the conversion from the local force on the bead to matrix stress. This gives a factor of ~2, ~3 and ~10 for collagen, fibrin, and Matrigel respectively. Note, to empirically find the conversion factor between force and stress by shifting, we use a comparable stiffening range of the microrheology and macrorheology data where they share a similar power-law behavior, as indicated by the solid symbols in the Supplementary Figure 5. Interestingly, we observe that $k_{nl}$ becomes increasingly stiffer at large force as indicated by the pink symbols. However, this high stress regime is not accessible in macrorheology.

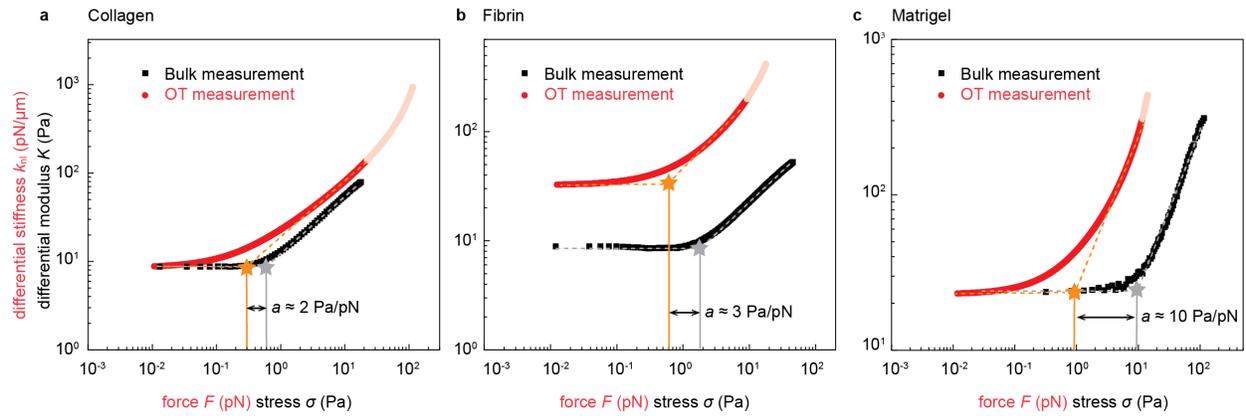

**Supplementary Figure 5:** Determination of conversion factors for NSIM between effective force and local matrix stress using active microrheology (Optical Tweezers) and bulk rheology, resulting in a factor of ~2, ~3 and ~10 for collagen, fibrin and Matrigel respectively.

**Supplementary Table 1:** The effective force $F_{eff}$ determined with NSIM using the data in Fig 4a

| Collagen | | Fibrin | | Matrigel | |
|---|---|---|---|---|---|
| $r$ (μm) | $F_{eff}$ (pN) | $r$ (μm) | $F_{eff}$ (pN) | $r$ (μm) | $F_{eff}$ (pN) |
| 2.6 | 114.50 | 3.3 | 10.50 | 3.4 | 14.28 |
| 5.5 | 19.68 | 8.0 | 2.00 | 6.9 | 7.60 |
| 10.5 | 5.40 | 15.5 | 0.26 | 14.8 | 1.47 |
| 17.2 | 1.43 | 24.4 | 0.03 | 24.5 | 0.59 |
| 28.9 | 0.44 | | | 33.1 | 0.43 |

## 2 Simulation methods and parameter choices

The simulation results presented in Figure 3 of the main text are obtained with a lattice model of nonlinear springs. In this model, nonlinear springs of unit rest length are arranged on the edges of a 3D face-centred cubic lattice, and the positions of the vertices thus represent the degrees of freedom of the system. To capture the stiffening behavior of the network, we choose a nonlinear force-extension relation for these springs as:

$$f(x) = \exp(\mu\, x) - 1 \tag{1}$$

which results in an exponential stress-strain relationship at macroscopic scales, consistent with experimental observations in collagen [32]. At large forces, $f$, this choice yields a power-law dependence of local differential stiffness on tension: $k_{\text{nl}} = df/dx \propto f$.

We simulate a large spherical region of the network of radius $R$ with fixed boundary conditions. The contractile cell is modelled by a rigid ellipsoid replacing a portion of the network in the center. All lattice nodes that are initially inside the ellipsoid are constrained to translate and rotate as a single rigid body, which can be affinely deformed to simulate contractile forces. Force and torque balance over the rigid body are ensured by including its bulk rotational and translational degrees of freedom to the energy minimization.

To capture network disorder, a fraction of the springs is randomly suppressed. This construction will result in substantial mechanical heterogeneities as well as non-affine network deformations, which are characteristic of fibrous networks [18]. However, a finite-size artefact occurs when a straight chain of springs remains that connects the cell at the center of the spherical network to the system's boundary. At large deformation, such straight lines tend to concentrate all the stress as they stiffen dramatically. Since 3D numerical simulations are limited in size, we choose to limit the length of such alignments to a given "fiber length" $L_f < R$. Specifically, for each line of aligned edges in the initial configuration, we randomly choose a starting point, and remove one spring every $L_f + 1$ edges, thereby ensuring that no straight line connecting the cell to the system's boundary remains.

Local stiffnesses are obtained by measuring the response to point forces, exerted on selected lattice nodes, and directed towards the center of the ellipse. The "remote" force-displacement curve in Figure 3b of the main text, which serves as a reference for NSIM, is obtained in a similar system with no contractile cell, and with the probe placed at the center of the network.

Local stress at a lattice node is defined as minus the dipole moment of the forces exerted by nearby nodes, divided by the unit cell volume. This construction is consistent with a macroscopic definition of stress.

Macrorheological calibration, as discussed in Figure 2 of the main text, is performed on a large system with periodic boundary conditions, in the absence of cells or probes. The network is stressed by affinely deforming the periodic boundary conditions vectors, in a Lees-Edwards fashion. The stress is evaluated using the discrete mean-stress theorem [46].

This system is simulated in C++14. Mechanical equilibrium is obtained by minimizing the total energy using the GNU Scientific Library [47] BFGS2 implementation of the Broyden-Fletcher-Goldfarb-Shanno algorithm. Data analysis and visualization are performed in Python2 using the SciPy [48], Matplotlib [49] and Mayavi2 [50] packages.

In Figure 3 of the main text, we use the following parameters: system radius $R = 25.5$, spring stiffness parameter $\mu = 50$, and fiber length $L_f = 10$. The cell has ellipse aspect ratio $5:1:1$, with its long axis with length 14.2 along the $(111)$ lattice direction. The cell is contracted by 50% along its long axis. We probe the linear stiffness at each of the 352 points that are present in the plane perpendicular to the $(1/\sqrt{7}, \sqrt{6/7}, 0)$ vector (which includes the cell's long axis) and located at a distance $> 5$ from the system's boundary to avoid edge effects. These points are displayed in Figure 3a of the main text. Local linear stiffness is measured by exerting a force strong enough to displace the probe node by $10^{-3}$, small enough to be consistently in the linear response regime, yet large enough to prevent numerical imprecision. The "remote" reference curve is obtained by measuring the displacement in response to forces ranging from $10^{-2}$ to $10^5$, in 22 logarithmic increments, and we used spline-fitting on the force-displacement curve in log-log space. This allows us to robustly evaluate the nonlinear differential stiffness $k_{\mathrm{nl}}(F)$, which we average over 20 realizations of the network. The macrorheological calibration system size is $40^3$, with isotropic strain $\gamma$ varying logarithmically from $10^{-5}$ to 0.46, and similar log-log spline-fitting is applied to macrorheological stress-strain curve. The consequences of this choice of geometry are discussed in Sec 3.4.

# 3 Robustness of NSIM

In this Section we provide further support for the robustness of our stress inference method in numerical simulations, by considering additional geometries, which are different from the case of stresses induced by an elliptic cell presented in Figure 3 of the main text.

## 3.1 Homogeneously prestressed systems

We first consider the case of a system where the stress is spatially homogeneous, and originates from the deformation of the boundary of the system. We consider a spherical system with fixed boundary conditions, with a deformation according to a strain:

$$\begin{pmatrix} \gamma & 0 & 0 \\ 0 & -\gamma/2 & 0 \\ 0 & 0 & -\gamma/2 \end{pmatrix} \qquad (2)$$

where the *x*-direction is also the force direction, which we take to be the (111) lattice direction. This geometry of rotationally symmetric shear strain corresponds to the strain in a linear elastic material with a spherically contractile cell.

We present the results for these simulations in Supplementary Figure **6**, showing excellent agreement between the inferred stress and its measured value, up to a proportionality factor of roughly $\sim 0.3$. This shows that NSIM accurately captures stress variations in a disordered networks, and that our calibration allows us to capture stresses within a factor of 2-3 depending on the geometry of the strain field. This accuracy is similar to other experimental errors such as the typical experiment-to-experiment variation of macrorheological protocols on biopolymer networks.

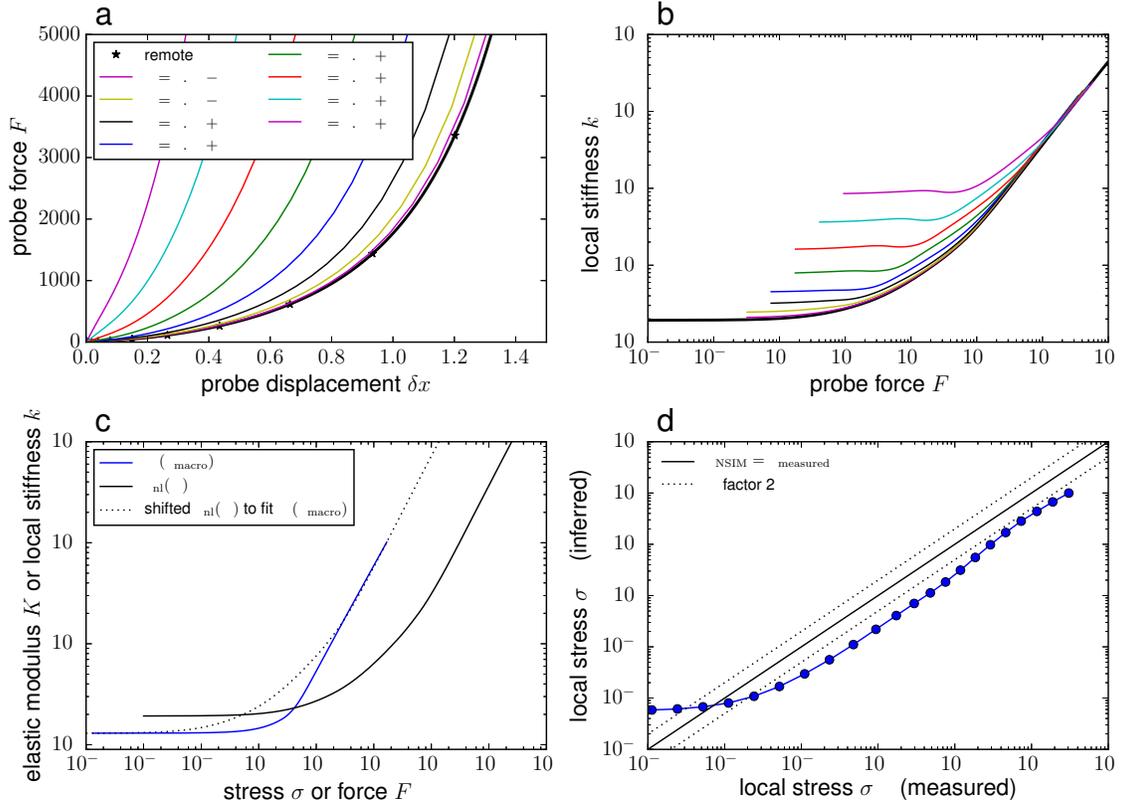

**Supplementary Figure 6**: Results for NSIM in a homogeneously prestrained network. **a**, Nonlinear force-displacement curves for varying pre-stress. **b**, Nonlinear stiffening curves (same colors as in **a**). **c**, Calibration using macrorheology data: the prefactor for NSIM is obtained by aligning the micro- and macrorheological curves, as illustrated in Figure 2 of the main text. **d**, Inferred local stress versus direct measurement, showing proportionality at large stress. Parameters for these simulations: $\mu = 100$, $L_f = 12$, $R = 15.5$.

## 3.2 Isotropically contractile cell

Next, we consider the case of a spherically symmetric contractile cell, modeled as a rigid sphere that shrinks isotropically (Supplementary Figure 7a). We perform NSIM on a 1000 random points in the network, whose distances to the cell center are uniformly distributed in log scale. These results are presented in Supplementary Figure 7c, showing excellent agreement between inferred and measured stress. The stress is long-ranged and decays as $r^{-2}$ with distance to the cell (Supplementary Figure 7e), consistent with previous results and simulations of elliptic cells presented in the main text.

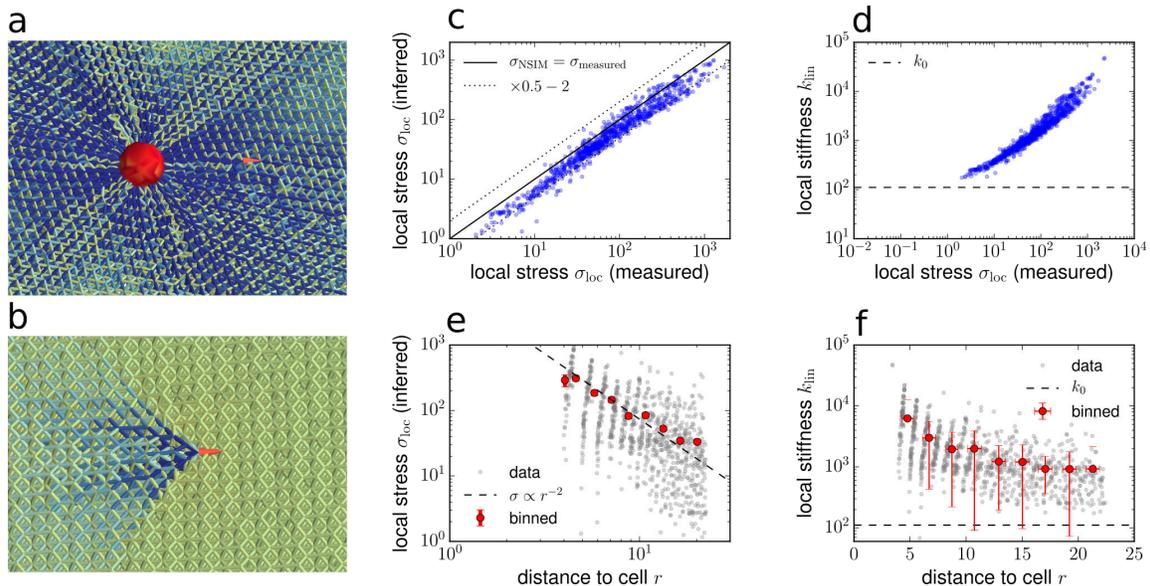

**Supplementary Figure 7**: NSIM in a spherically symmetric stress field. **a,** Here the cell model consists of a rigid sphere of radius $R_{\text{cell}} = 5.1$ that shrinks by 55% in a spherical system of radius $R = 25.5$, with fixed boundary conditions. Network parameters: $L_f = 22$, $\mu = 50$. **b,** The "remote" reference curve is obtained in an unstressed network. **c,** Stress inferred with NSIM *vs* measured stress, for a 1000 random points in the network. The agreement between the two measures is good over three decades of stress. **d,** Linear stiffness *vs* stress. **e,** Stress decay away from the cell. While the signal-to-noise ratio is high due to the mechanical disorder intrinsic to our modelled networks, binning and averaging yield results consistent with rope-like force transmission. **f,** Decay of the local stiffness away from the cell.

## 3.3 NSIM captures local stress fluctuations

In this section we show that NSIM captures not only the average stress, but also local stress fluctuations due to heterogeneity in the local mechanical properties. To demonstrate this, we consider our simulations in the case of an isotropically contractile cell (Sec 3.2) and analyze the correlations between three data sets: distances to cell $r$, local linear stiffness $k_{\text{lin}}$, and local stress $\sigma_{\text{loc}}$ (obtained by direct measurement). We find the following Pearson linear correlation coefficients $\rho$ (for log-log correlations):

$$\rho(k_{\text{lin}}, r) = -0.63$$
$$\rho(\sigma_{\text{loc}}, r) = 0.56$$
$$\rho(k_{\text{lin}}, \sigma_{\text{loc}}) = 0.97$$

In other words, both $k_{\text{lin}}$ and $\sigma_{\text{loc}}$ fluctuate strongly and are not determined by the distance to the cell, but stress fluctuations are strongly correlated with local stiffness fluctuations.

Performing a log-log linear regression of the measured local stress $\sigma_{\text{loc}}$ with that obtained using NSIM, we find that the best power-law fit has exponent 1.04 and prefactor 0.567 (the prefactor here is with isotropic prestrain for the calibration; see discussion in sec 3.4). This is consistent with the ideal case for NSIM where the exponent would be 1, showing that NSIM also captures stress fluctuations due to mechanical heterogeneity.

### 3.4 Macrorheological geometry and NSIM prefactor

Calibration is done with a macrorheology simulation, measuring the stress response to a bulk strain with a system size of $40^3$ with periodic boundary conditions. Tensorial stresses are measured using the mean stress theorem.

If $\hat{\gamma} = \gamma \hat{e}_\gamma$ is the strain matrix, we define the scalar stress as $\sigma = \text{Tr}(\hat{\gamma} \cdot \hat{\sigma})/||\hat{\gamma}||$ and the differential modulus as:

$$K = \frac{\partial \sigma}{\partial \gamma}$$

using a logarithmic spline interpolation to differentiate.

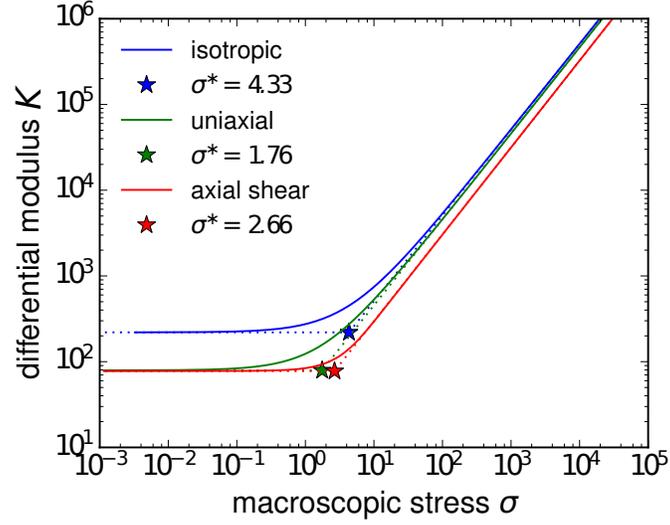

**Supplementary Figure 8**: Macrorheology curves for calibration. Different strain geometries lead to different values of $\sigma^*$, which impacts the final value of the force-stress conversion factor. Empirically, we find that an isotropic dilation (blue curve) gives the most accurate result for the conversion factor. The green curve shows the case of a uniaxial dilation, and the red curve corresponds to uniaxial dilation with compensating compression in the other two directions to ensure volume conservation at linear level. A spherical contractile cell in linear elasticity corresponds to the latter case.

# 4  Proof of NSIM within a minimal 1D model

In the main text we mentioned that the NSIM correspondence between local force and local stress at equivalent stiffness becomes exact in the case of power-law stiffening of the network. To establish this result, we employ a minimal model, consisting of two nonlinear springs describing the network response.

## 4.1  Model

Our minimal model (Supplementary Figure **9**) consists of two nonlinear springs in series, with unit rest length and a generic force-extension relation $\frac{1}{2}f(\delta)$ (the 1/2 being for convenience of notation). The network prestress is modeled by a displacement of the two boundary points by $\gamma$ (fixed strain), resulting in a stress

$$\sigma(\gamma) = f(\gamma) \qquad (3)$$

at the central point. The force exerted by optical tweezers results in a displacement of the central point of a distance $x$, corresponding to a force

$$F(x,\gamma) = \frac{f(\gamma+x) - f(\gamma-x)}{2} \qquad (4)$$

In particular, in the absence of prestress, we define the reference force-displacement curve for the probe,

$$F_{\text{ref}}(x) = F(x,\gamma=0) = \frac{f(x) - f(-x)}{2} = \tilde{f}(x) \qquad (5)$$

where $\tilde{f}$ is the odd part of $f$.

As discussed in the main text, the experimentally measurable quantities are $F(x,\gamma)$ for an unknown prestress. We define as before the linear stiffness in the pre-strained state:

$$k_{\text{lin}}(\gamma) = \frac{\partial F}{\partial x}(x=0,\gamma) = f'(\gamma) \qquad (6)$$

and the nonlinear differential stiffness in the absence of pre-stress:

$$k_{\text{nl}}(x) = \frac{\partial F}{\partial x}(x,\gamma=0) = \frac{f'(x) + f'(-x)}{2} = \tilde{f}'(x) \qquad (7)$$

## 4.2  Mathematical formulation of NSIM correspondence

The idea of Nonlinear Stress Inference Microscopy is that the network's stiffness depends on its stress state (*i.e.*, $f$ is nonlinear) such that we can establish a correspondence between the force-induced stiffening curve $k_{\text{nl}}$ and the stress-induced stiffening curve $k_{\text{lin}}$.

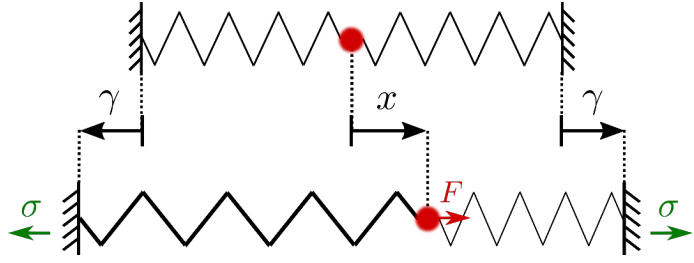

**Supplementary Figure 9**: A minimal model for our microrheological system. Two nonlinear springs describe the network surrounding a bead. Starting from an unstressed reference configuration (top), the cell-generated prestress, $\sigma$, induces a symmetric elongation of the two springs by $\gamma$, and the force $F$ exerted by the optical tweezer results in a displacement $x$ of the bead.

Specifically, to each pre-strain $\gamma$ we associate an *effective displacement* $x_{\text{eff}}(\gamma)$ corresponding to the same differential stiffness, such that

$$k_{\text{lin}}(\gamma) = k_{\text{nl}}\left(x_{\text{eff}}(\gamma)\right) \tag{8}$$

hence

$$x_{\text{eff}}(\gamma) = k_{\text{nl}}^{-1}\left(k_{\text{lin}}(\gamma)\right) \tag{9}$$

We are interested, however, in mechanical quantities $F$ and $\sigma$ rather than in the associate geometrical quantities $x$ and $\gamma$. We thus define the effective local force $F_{\text{eff}}(\sigma)$ that corresponds to the same differential stiffness as the local stress $\sigma$:

$$F_{\text{eff}}(\sigma) = F_{\text{ref}}\left(k_{\text{nl}}^{-1}\left(k_{\text{lin}}\left(\gamma(\sigma)\right)\right)\right) \tag{10}$$

Our stress microscopy technique requires that $F_{\text{eff}}(\sigma) \propto \sigma$; we now study the conditions under which this identity holds.

Mathematically, in terms of the generic nonlinear force-extension relation of the springs, we have:

- 
$$\sigma(\gamma) = f(\gamma) \tag{11}$$

- 
$$k_{\text{lin}}(\gamma) = f'(\gamma) \tag{12}$$

- 
$$F_{\text{ref}}(x) = \frac{f(x) - f(-x)}{2} = \tilde{f}(x) \tag{13}$$

where $\tilde{f}$ is the odd part of $f$.

- 

$$k_{\text{nl}}(x) = \tilde{f}'(x) \tag{14}$$

therefore:

$$F_{\text{eff}}(\sigma) = \left[\tilde{f} \circ \left(\tilde{f}'\right)^{-1} \circ f' \circ f^{-1}\right](\sigma) \tag{15}$$

Schematically:

$$\sigma \xmapsto{f^{-1}} \gamma \xmapsto{f'} k_{\text{lin}} \sim k_{\text{nl}} \xmapsto{\left(\tilde{f}'\right)^{-1}} x_{\text{eff}} \xmapsto{\tilde{f}} F_{\text{eff}} \tag{16}$$

### 4.3 Conditions for NSIM

The question we need to address is: how does the composite function $F_{\text{eff}} = \tilde{f} \circ \left(\tilde{f}'\right)^{-1} \circ f' \circ f^{-1}$ compare to the identity function? When this is the case, we will be ensured to have: $F_{\text{eff}}(\sigma) \propto \sigma$, which is the requirement for NSIM.

Note first that in order to be able to infer an effective force $F_{\text{eff}}$, the values taken by $k_{\text{nl}}$ and $k_{\text{lin}}$ should be the same. In particular, this implies that $f$ should have a *nonlinear odd* component, such as a cubic or quintic term. Otherwise, there is no stiffening under displacement of the probe (as the weakening of one side exactly compensates the stiffening of the other), and thus no inference is possible as $\tilde{f}$ is not invertible. In contrast, if $f$ is odd, $F_{\text{eff}}(\sigma) = \sigma$ for all values of $\sigma$; however this case corresponds to a material for which the nonlinear response under compression and tension is the same, which does not correspond to any physical situation. Also $\tilde{f}'$ is even by construction, and thus invertible only on $\mathbb{R}_+$, *i.e.*, probe forces $F$ and $-F$ are indistinguishable.

#### 4.3.1 Weakly nonlinear regime

Next we consider a generic case where all terms of the Taylor expansion of $f$ are nonzero:

$$f(\delta) = \delta + \frac{a_2}{2}\delta^2 + \frac{a_3}{3}\delta^3 + \ldots \tag{17}$$

where the first coefficient is set to unity (which gives the stiffness scale, *i.e.* $k_0 = 1$). In this case, we have, to first nonlinear order:

$$f'(\delta) = 1 + a_2\delta + \ldots \tag{18}$$
$$\tilde{f}(\delta) = \delta + \frac{a_3}{3}\delta^3 + \ldots \tag{19}$$
$$\tilde{f}'(\delta) = 1 + a_3\delta^2 + \ldots \tag{20}$$

thus

$$f^{-1}(t) = t - \frac{a_2}{2}t^2 \tag{21}$$
$$\left(\tilde{f}'\right)^{-1}(k) = \sqrt{\frac{k-1}{a_3}} + \ldots \tag{22}$$

and so
$$F_{\text{eff}}(\sigma) = \left[\tilde{f} \circ \left(\tilde{f}'\right)^{-1} \circ f' \circ f^{-1}\right](\sigma) = \sqrt{\sigma} + \left(\frac{a_3}{3} - \frac{a_2}{4}\right)\sigma^{3/2} + \ldots \quad (23)$$

Hence in the weakly nonlinear regime $F_{\text{eff}}(\sigma) \sim \sqrt{\sigma}$, precluding an accurate inference of stress by this method. This is due to the difference in symmetry between force and stress, as discussed in Figure 2 of the main text.

### 4.3.2 Strongly nonlinear regime

We now examine the conditions for proportionality between stress and effective force in the strongly nonlinear regime. We make the following assumptions regarding the stiffening function $f$ characteristic of the material:

- Diverging stiffness:
$$\lim_{x \to x_\infty} f'(x) = \infty \quad (24)$$

  where $x_\infty$ stands for either a finite maximum elongation (Worm-Like-Chain-like divergence) or infinity. This assumption excludes linear and strain-softening materials for instance.

- Compression-extension ratio:
$$\lim_{x \to x_\infty} f(x)/f(-x) = -c \quad (25)$$

  where $0 \leq c < \infty$. The number $c$ is a characteristic of the material: $c = 0$ for a material which buckles and yields under compression, and $c = 1$ for an odd force-extension relation. We also define $q = (1+c)/2$, such that:
$$\tilde{f}(x) \underset{x \to x_\infty}{\sim} q f(x) \quad (26)$$

  with $q = 1/2$ for a buckling material.

Under these assumptions, we now derive a functional relation that the force extension curve $f$ needs to obey to exactly satisfy NSIM conditions in the strongly nonlinear regime. First we define $S(k) = f\left(f'^{-1}(k)\right)$ and $\tilde{S}(k) = \tilde{f}\left(\tilde{f}'^{-1}(k)\right)$, such that:
$$\sigma = S(k_{\text{lin}}) \quad (27)$$

and
$$F = \tilde{S}(k_{\text{nl}}) \quad (28)$$

Hence we can rewrite Eq. 23 as
$$F_{\text{eff}}(\sigma) = \tilde{S}\left(S^{-1}(\sigma)\right) \quad (29)$$

The requirement for our NSIM correspondence is that $F_{\text{eff}}(\sigma) \propto \sigma$, which translates to $\tilde{S}\left(S^{-1}(\sigma)\right) = b\sigma$ for some stress-force proportionality constant $b$. For this proportionality to hold in the strongly nonlinear asymptotic regime $k \to \infty$ limit (under the assumption that the network stiffens under tension, *i.e.* Eq. (24)), we need to have

$$\frac{S(k)}{\tilde{S}(k)} \xrightarrow[k \to \infty]{} b \quad ; \quad 0 < b < \infty \tag{30}$$

Equation (26) implies that

$$\tilde{f}'^{-1}(k) \underset{k \to \infty}{\sim} f'^{-1}(k/q) \tag{31}$$

Combining this with Eq. 30, we get

$$\tilde{S}(k) \underset{k \to \infty}{\sim} qS(k/q) \tag{32}$$

The condition that $S$ and $\tilde{S}$ be proportional (Eq. (30)) thus implies:

$$\boxed{S(k) \underset{k \to \infty}{\sim} bqS(k/q)} \tag{33}$$

Of course, if $q = 1$ (asymptotically odd function), this equation only implies $b = 1$, which we already knew (as discussed at the start of Sec. 4.3). For any other value of $q$, including the case of a buckling material, Eq. (33) is highly constraining: Assuming a physically reasonable behavior for the material[1], this functional relationship indeed implies a power-law behavior prescribed by the value of the proportionality constant $b$:

$$S \underset{k \to \infty}{\sim} Ak^\alpha \tag{34}$$

with

$$\alpha = 1 + \frac{\log b}{\log q} \tag{35}$$

Note that the values of $q$ and $\alpha$, both of which are physically measurable and meaningful, set the value of the constant $b$.

### 4.3.3 Consequences on $f$

Recalling the definition of $S(k) = f\left(f'^{-1}(k)\right)$ (Eq. (27)), the power law form (rewriting Eq. (34) in terms of $f^{-1}$) implies that :

$$f'^{-1}(k) \underset{k \to \infty}{\sim} f^{-1}(Ak^\alpha) \tag{36}$$

We invert this equation to arrive at

$$f'(x) \underset{x \to x_\infty}{\sim} \left[\frac{1}{A}f(x)\right]^{\frac{1}{\alpha}} \tag{37}$$

---

[1] More precisely Eq. (33) implies that $S(k) = k^\alpha \phi(\log k)$, where $\phi$ is a generic $\log(q)$-periodic function. It is thus natural to assume that $\phi$ is a constant, the alternative being that the stiffness curve shows some kind of non-trivial self-similar structure.

This differential equation on $f$ has the following solution, depending on the value of $\alpha$:

$$f(x) \sim \begin{cases} x^{\frac{\alpha}{1-\alpha}} & 0 < \alpha < 1 \\ \exp(\mu x) & \alpha = 1 \\ (x_\infty - x)^{-\frac{\alpha}{\alpha-1}} & \alpha > 1 \end{cases} \tag{38}$$

These solutions include standard models for force-extension relations of individual biopolymers, such as the Worm-Like-Chain model ($\alpha = 2$) or exponential stiffening ($\alpha = 1$ as in the numerical simulations presented in the main text).

In summary, we have shown that the effective force $F_{\text{eff}}$ becomes exactly proportional to the local stress $\sigma_{\text{loc}}$ in the strongly nonlinear regime if, and only if, the stiffness-tension relationship of the springs asymptotically behaves as a power-law.